\documentclass{llncs}
\usepackage{llncsdoc}

\usepackage{graphicx}
\usepackage{hyperref}

\usepackage{amsmath}
\usepackage[algo2e]{algorithm2e}
\usepackage{algorithm}
\usepackage{algpseudocode}
\usepackage{algorithmicx}

\usepackage{listings}

\usepackage{csquotes}
\usepackage{subcaption}

\begin{document}

\title{
Stay On-Topic: 
Generating \\Context-specific 
Fake Restaurant Reviews
}



\author{Mika Juuti\inst{1} \and Bo Sun\inst{2} \and
Tatsuya Mori\inst{3} \and N. Asokan\inst{1}}

\institute{Aalto University, Finland\\
\email{mika.juuti@aalto.fi}, \email{asokan@acm.org},\\
\and
National Institute of Information and Communications Technology and Waseda University, Japan \\
\email{bo\_sun@nict.go.jp\footnote{Partially completed during his PhD course at Waseda University.}}
\and
Waseda University and RIKEN Center for Advanced Intelligence Project, Japan\\
\email{mori@nsl.cs.waseda.ac.jp}
}

\maketitle     

\begin{abstract}
Automatically generated fake restaurant reviews are a threat to online review systems. 
Recent research has shown that users have difficulties in detecting machine-generated 
fake reviews hiding among real restaurant reviews. 
The method used in this work (\emph{char-LSTM}) has one drawback: it has difficulties staying in \emph{context},
i.e. when it generates a review for specific target entity, the resulting review may contain phrases that are unrelated to the target, thus increasing its detectability.
In this work, we present and evaluate a more sophisticated
technique based on neural machine translation (NMT) with 
which we can generate reviews that \emph{stay on-topic}.
We test multiple variants of our technique using
native English speakers on Amazon Mechanical Turk. 
We demonstrate that reviews generated by the best variant have almost optimal undetectability (class-averaged F-score 47\%). 
We conduct a user study with experienced users and show that our method evades detection more frequently compared to the state-of-the-art (average evasion $3.2/4$ vs $1.5/4$) with statistical significance, at level $\alpha = 1\%$ (Section~\ref{sec:comparison}).
We develop very effective detection tools and reach average F-score of $97\%$ in classifying these. 
Although fake reviews are very effective in fooling people, effective automatic detection is still feasible.

\end{abstract}

\section{Introduction}

Automatically generated fake reviews have only recently become natural enough to
fool human readers. Yao et al. \cite{yao2017automated} use a deep neural network (a so-called 2-layer  LSTM\cite{murphy2012machine}) to generate fake reviews, and concluded that these fake reviews look sufficiently genuine to fool native English speakers. 
They train their model using real restaurant reviews from yelp.com \cite{challenge2013yelp}.
Once trained, the model is used to generate reviews character-by-character. 
Due to the generation methodology, it cannot be easily targeted for a specific \emph{context} (meaningful side information). 
Consequently, the review generation process may stray \emph{off-topic}. For instance, when generating a review for a Japanese restaurant in Las Vegas, the review generation process may include references to an Italian restaurant in Baltimore. The authors of \cite{yao2017automated} apply a post-processing step (\textit{customization}), which replaces food-related words with
more suitable ones (sampled from the targeted restaurant). The word replacement strategy has drawbacks: it can miss certain words and replace others independent of their surrounding words, which may alert savvy readers. 
As an example: when we applied the customization technique described in \cite{yao2017automated} to a review for a Japanese restaurant it changed the snippet \textit{garlic knots for \underline{breakfast}} with \textit{garlic knots for \underline{sushi}}).

We propose a methodology based on neural machine translation (NMT) that improves the generation process by defining a context for the each generated fake review. 
Our context is a clear-text sequence of: the review rating, restaurant name, city, state and food tags (e.g. Japanese, Italian). 
We show that our technique generates review that \emph{stay on topic}. 
We can instantiate our basic technique into several variants. 
We vet them on Amazon Mechanical Turk and find that native English speakers are very poor at recognizing our fake generated reviews. 
For one variant, 
the participants' performance is close to random: the class-averaged F-score of detection is $47\%$ (whereas random would be $42\%$ given the 1:6 imbalance in the test). 
Via a user study with experienced, highly educated participants, we compare this variant (which we will henceforth refer to as \emph{NMT-Fake* reviews}) with fake reviews generated using the char-LSTM-based technique from \cite{yao2017automated}.

We demonstrate that NMT-Fake* reviews constitute a new category of fake reviews that \emph{cannot be detected} by classifiers trained only using previously known categories of fake reviews \cite{yao2017automated,mukherjee2013yelp,rayana2015collective}. Therefore, NMT-Fake* reviews may go undetected in existing online review sites.
To meet this challenge, we develop an effective classifier that detects NMT-Fake* reviews effectively (97\% F-score). Our main contributions are:
\begin{itemize}
\item We present a novel method for creating machine-generated fake user reviews that \textbf{generates content based on specific context}: venue name, user rating, city etc (Sections \ref{sec:model} to \ref{sec:generating}). 
We demonstrate that our model can be trained faster (90\% reduction in training time compared to \cite{yao2017automated}, Section~\ref{sec:generating}) and resulting NMT-Fake* reviews are \textbf{highly effective in fooling native English speakers} (class-averaged F-score 47\%, Section~\ref{sec:amt}).
\item We \textbf{reproduce} a previously proposed \textbf{fake review generation method} \cite{yao2017automated} (Section~\ref{sec:repl}) and show that NMT-Fake* reviews are \textbf{statistically different} from previous fake reviews, and that classifiers trained on previous fake review types do \textbf{not detect} NMT-Fake* reviews (Section~\ref{sec:automated}). 
\item We compare NMT-Fake* reviews with char-LSTM reviews in a user study. 
We show that our reviews are \textbf{significantly better at evading detection} with statistical significance ($\alpha = 1\%$) (Section~\ref{sec:comparison}). 
\item We develop \textbf{highly efficient statistical detection tools} to recognize NMT-Fake* reviews with 97\% F-score (Section~\ref{sec:detection}). We plan to share the implementation of our detector and generative model with other researchers to facilitate transparency and reproducibility.
\end{itemize}

\section{Background}

\noindent{\bf Fake reviews}
User-generated content \cite{o2008user} is an integral part of the contemporary user experience on the web. 
Sites like \emph{tripadvisor.com}, \emph{yelp.com} and \emph{Google Play} use user-written reviews to provide rich information that helps other users choose where to spend money and time. User reviews are used for rating services or products, and for providing qualitative opinions. User reviews and ratings may be used to rank services in recommendations. Ratings have an affect on the outwards appearance. Already 8 years ago, researchers estimated that a one-star rating increase affects the business revenue by 5 -- 9\% on yelp.com \cite{luca2010reviews}.

Due to monetary impact of user-generated content, some businesses have relied on so-called crowd-turfing \emph{agents} \cite{wang2012serf} that promise to deliver positive ratings written by \emph{workers} to a \emph{customer} in exchange for a monetary compensation. Crowd-turfing ethics are complicated. For example, Amazon community guidelines prohibit buying content relating to promotions, but the act of writing fabricated content is not considered illegal, nor is matching workers to customers \cite{rinta2017understanding}. 
Year 2015, approximately 20\% of online reviews on yelp.com were suspected of being fake \cite{luca2016fake}.

Nowadays, user-generated review sites like yelp.com use filters and fraudulent review detection techniques. 
These factors have resulted in an increase in the requirements of crowd-turfed reviews provided to review sites, which in turn has led to an increase in the cost of high-quality review. Due to the cost increase, researchers hypothesize the existence of neural network-generated fake reviews. These neural-network-based fake reviews are statistically different from human-written fake reviews, and are not caught by classifiers trained on these \cite{yao2017automated}. 

Detecting fake reviews can either be done on an individual level or as a system-wide detection tool (i.e. regulation). Detecting fake online content on a personal level requires knowledge and skills in critical reading. 
In 2017, the National Literacy Trust assessed that young people in the UK do not have the skillset to differentiate fake news from real news \cite{national2017commission}.
For example, 20\% of children that use online news sites in age group 12-15 believe that all information on news sites are true. 

\noindent{\bf Neural Networks}
Neural networks are function compositions that map input data through $k$ subsequent layers:

\begin{equation}
F(x) = f_k \circ f_{k-1} \circ \dots \circ f_{2} \circ f_{1} \circ x,
\end{equation}

where the functions $f_k$ are typically non-linear and chosen by experts partly for known good performance on datasets and partly for simplicity of computational evaluation. Language models (LMs) \cite{jurafsky2014speech} are generative probability distributions that assign probabilities to sequences of tokens ($t_i$):

\begin{equation}
p(t_k | t_{<k}) = p(t_k | t_{k-1}, t_{k-2}, \dots, t_{2}, t_{1}),
\end{equation}

such that the language model can be used to predict how likely a specific token at time step $k$ is, based on the $k-1$ previous tokens. Tokens are typically either words or characters.

For decades, deep neural networks were thought to be computationally too difficult to train. However, advances in optimization, hardware and the availability of frameworks have shown otherwise \cite{murphy2012machine}, \cite{kingma2014adam}. 
Neural language models (NLMs) have been one of the promising application areas. 
NLMs are typically various forms of recurrent neural networks (RNNs), which pass through the data sequentially and maintain a memory representation of the past tokens with a hidden context vector. There are many RNN architectures that focus on different ways of updating and maintaining context vectors: Long Short-Term Memory units (LSTM) and Gated Recurrent Units (GRUs) are perhaps most popular. Neural LMs have been used for free-form text generation. 
In certain application areas, the quality has been high enough to sometimes fool human readers \cite{yao2017automated}.

Encoder-decoder (seq2seq) models \cite{cho2014learning} are architectures of stacked RNNs, which have the ability to generate output sequences based on input sequences. The encoder network reads in a sequence of tokens, and passes it to a decoder network (a LM). In contrast to simpler NLMs, encoder-decoder networks have the ability to use additional context for generating text, which enables more accurate generation of text. 
Encoder-decoder models are integral in \emph{Neural Machine Translation} (\emph{NMT}) \cite{klein2017opennmt}, where the task is to translate a source text from one language to another language. NMT models additionally use beam search strategies to heuristically search the set of possible translations. Training datasets are parallel corpora; large sets of paired sentences in the source and target languages. 
The application of NMT techniques for online machine translation has significantly improved the quality of translations, bringing it closer to human performance \cite{wu2016google}.

Neural machine translation models are efficient at mapping one expression to another (one-to-one mapping). Researchers have evaluated these models for conversation generation \cite{mei2017coherent}, with mixed results. Some researchers attribute poor performance to the use of the negative log likelihood cost function during training, which emphasizes generation of high-confidence phrases rather than diverse phrases \cite{li2016diversity}. The results are often generic text, which lacks variation. 
Li et al. have suggested various augmentations to this, among others suppressing typical responses in the decoder language model to promote response diversity \cite{li2016diversity}.

\section{System Model}
\graphicspath{ {figures/}}
We discuss the attack model, our generative machine learning method and controlling the generative process in this section.

\subsection{Attack Model}



Wang et al. \cite{wang2012serf} described a model of crowd-turfing attacks consisting of three entities: \textbf{customers} who desire to have fake reviews for a particular target (e.g. their restaurant) on a particular platform (e.g. Yelp), \textbf{agents} who offer fake review services to customers, and \textbf{workers} who are orchestrated by the agent to compose and post fake reviews.

Automated crowd-turfing attacks (ACA) replace workers by a \textbf{generative model}. This has several benefits including better economy and scalability (human workers are more expensive and slower) and reduced detectability (agent can better control the rate at which fake reviews are generated and posted). 

We assume that the agent has access to public reviews on the review platform, by which it can train its generative model. We also assume that it is easy for the agent to create a large number of accounts on the review platform so that account-based detection or rate-limiting techniques are ineffective against fake reviews.

The quality of the generative model plays a crucial role in the attack. Yao et al. \cite{yao2017automated} propose the use of a character-based LSTM as base for generative model. LSTMs are not conditioned to generate reviews for a specific target \cite{murphy2012machine}, and may mix-up concepts from different \emph{contexts} during free-form generation. Mixing contextually separate words is one of the key criteria that humans use to identify fake reviews. 
These may result in violations of known indicators for fake content \cite{rubin2006assessing}. For example, the review content may not match prior expectations nor the information need that the reader has. 
We improve the attack model by considering a more capable generative model that produces more appropriate reviews: a neural machine translation (NMT) model.


\subsection{Generative Model}
\label{sec:model}

\subsubsection{Architecture}
We propose the use of NMT models for fake review generation. The method has several benefits: 1) the ability to \emph{learn} how to associate context (keywords) to reviews, 2) \emph{fast} training time, and 3) a high-degree of \emph{customization} during production time, e.g. introduction of specific waiter or food items names into reviews.

NMT models are constructions of stacked recurrent neural networks (RNNs). They include an \emph{encoder} network and a \emph{decoder} network, which are jointly optimized to produce a \emph{translation} of one sequence to another. The encoder  rolls over the input data in sequence and produces \emph{one} $n$-dimensional context vector representation for the sentence. The decoder then generates output sequences based on the embedding vector and an \emph{attention module}, which is taught to associate output words with certain input words. The generation typically continues until a specific \emph{EOS} (end of sentence) token is encountered. The review length can be controlled in many ways, e.g. by setting the probability of generating the EOS token to zero until the required length is reached.

NMT models often also include a beam search \cite{klein2017opennmt}, which generates several hypotheses and chooses the best ones amongst them. In our work, we use the greedy beam search technique. We forgo the use of additional beam searches as we found that the quality of the output was already adequate and the translation phase time consumption increases linearly for each beam used. 

\subsubsection{Dataset}

We use the Yelp Challenge dataset \cite{challenge2013yelp} for our fake review generation. The dataset (Aug 2017) contains 2.9 million 1 \textendash 5 star restaurant reviews. We treat all reviews as genuine human-written reviews for the purpose of this work, since wide-scale deployment of machine-generated review attacks are not yet reported (Sep 2017) \cite{zhao2017news}.
As preprocessing, we remove non-printable (non-ASCII) characters and excessive white-space. We separate punctuation from words. 
We reserve 15,000 reviews for validation and 3,000 for testing, and the rest we use for training.
NMT models require a parallel corpus of source and target sentences, i.e. a large set of (source, target)-pairs. 
We set up a parallel corpus by constructing (context, review)-pairs from the dataset.
Next, we describe how we created our input context.

\subsubsection{Context}

The Yelp Challenge dataset includes metadata about restaurants, including their names, food tags, cities and states these restaurants are located in. For each restaurant review, we fetch this metadata and use it as our input context in the NMT model. The corresponding restaurant review is similarly set as the target sentence. This method produced 2.9 million pairs of sentences in our parallel corpus. We show one example of the parallel training corpus in Example 1 below:

\noindent\begin{verbatim} Example 1.
5 Public House Las Vegas NV Gastropubs Restaurants > Excellent
food and service . Pricey , but well worth it . I would recommend 
the bone marrow and sampler platter for appetizers . \end{verbatim}


\noindent The order {\textbf{[rating name city state tags]}} is kept constant. 
Training the model conditions it to associate certain sequences of words in the input sentence with others in the output. 

\subsubsection{Training Settings}

We train our NMT model on a commodity PC with a i7-4790k CPU (4.00GHz), with 32GB RAM and one NVidia GeForce GTX 980 GPU. Our system can process approximately 1,300 \textendash 1,500 source tokens/s and approximately 5,730 \textendash 5,830 output tokens/s. Training one epoch takes in average 72 minutes. The model is trained for 8 epochs, i.e. over night. We call fake review generated by this model \emph{NMT-Fake reviews}. We only need to train one model to produce reviews of different ratings.
We use the training settings: adam optimizer \cite{kingma2014adam} with the suggested learning rate 0.001 \cite{klein2017opennmt}. For most parts, parameters are at their default values. Notably, the maximum sentence length of input and output is 50 tokens by default. 
We leverage the framework openNMT-py \cite{klein2017opennmt} to teach the our NMT model. 
We list used openNMT-py commands in Appendix Table~\ref{table:openNMT-py_commands}. 

\begin{figure}[t]
\begin{center}
  \begin{tabular}{ | l |   }
    \hline
Example 2.    Greedy NMT \\
Great food, \underline{great} service, \underline{great} \textit{\textit{beer selection}}. I had the \textit{Gastropubs burger} and it
\\
was delicious. The \underline{\textit{beer selection}} was also \underline{great}. \\
\\
Example 3. NMT-Fake* \\
I love this restaurant. Great food, great service. It's \textit{a little pricy} but worth\\
it for the \textit{quality} of the \textit{beer} and atmosphere you can see in \textit{Vegas}
\\
  \hline
  \end{tabular}
  \label{table:output_comparison}
\end{center}
\caption{Na\"{i}ve text generation with NMT vs. generation using our NTM model. Repetitive patterns are \underline{underlined}. Contextual words are \emph{italicized}. Both examples here are generated based on the context given in Example~1.}
\label{fig:comparison}
\end{figure}

\subsection{Controlling generation of fake reviews}
\label{sec:generating}

Greedy NMT beam searches are practical in many NMT cases. However, the results are simply repetitive, when naively applied to fake review generation (See Example~2 in Figure~\ref{fig:comparison}).
The NMT model produces many \emph{high-confidence} word predictions, which are repetitive and obviously fake. We calculated that in fact, 43\% of the generated sentences started with the phrase ``Great food''. The lack of diversity in greedy use of NMTs for text generation is clear.

\begin{algorithm}[!b]
 \KwData{Desired review context $C_\mathrm{input}$ (given as cleartext), NMT model}
 \KwResult{Generated review $out$ for input context $C_\mathrm{input}$}
set $b=0.3$, $\lambda=-5$, $\alpha=\frac{2}{3}$,  $p_\mathrm{typo}$, $p_\mathrm{spell}$ \\
$\log p \leftarrow \text{NMT.decode(NMT.encode(}C_\mathrm{input}\text{))}$ \\
out $\leftarrow$ [~] \\
$i \leftarrow 0$ \\
$\log p \leftarrow \text{Augment}(\log p$, $b$, $\lambda$, $1$, $[~]$, 0)~~~~~~~~~~~~~~~ |~random penalty~\\ 
\While{$i=0$ or $o_i$ not EOS}{ 
$\log \Tilde{p} \leftarrow \text{Augment}(\log p$, $b$, $\lambda$, $\alpha$, $o_i$, $i$)~~~~~~~~~~~ |~start \& memory penalty~\\ 
$o_i \leftarrow$ \text{NMT.beam}($\log \Tilde{p}$, out) \\
out.append($o_i$) \\
$i \leftarrow i+1$
}
\text{return}~$\text{Obfuscate}$(out,~$p_\mathrm{typo}$,~$p_\mathrm{spell}$)
\caption{Generation of NMT-Fake* reviews.}
\label{alg:base}
\end{algorithm}

In this work, we describe how we succeeded in creating more diverse and less repetitive generated reviews, such as Example 3 in Figure~\ref{fig:comparison}. 
We outline pseudocode for our methodology of generating fake reviews in Algorithm~\ref{alg:base}. There are several parameters in our algorithm. 
The details of the algorithm will be shown later. 
We modify the openNMT-py translation phase by changing log-probabilities before passing them to the beam search.
We notice that reviews generated with openNMT-py contain almost no language errors. As an optional post-processing step, we obfuscate reviews by introducing natural typos/misspellings randomly. In the next sections, we describe how we succeeded in generating more natural sentences from our NMT model, i.e. generating reviews like Example~3 instead of reviews like Example~2.

\subsubsection{Variation in word content}

Example 2 in Figure~\ref{fig:comparison} repeats commonly occurring words given for a specific context (e.g. \textit{great, food, service, beer, selection, burger} for Example~1). Generic review generation can be avoided by decreasing probabilities (log-likelihoods \cite{murphy2012machine}) of the generators LM, the decoder. 
We constrain the generation of sentences by randomly \emph{imposing penalties to words}. 
We tried several forms of added randomness, and found that adding constant penalties to a \emph{random subset} of the target words resulted in the most natural sentence flow. We call these penalties \emph{Bernoulli penalties}, since the random variables are chosen as either 1 or 0 (on or off).

\paragraph{Bernoulli penalties to language model}
To avoid generic sentences components, we augment the default language model $p(\cdot)$ of the decoder by

\begin{equation}
\log \Tilde{p}(t_k) =  \log p(t_k | t_i, \dots, t_1) + \lambda q,
\end{equation}

where $q \in R^{V}$ is a vector of Bernoulli-distributed random values that obtain values $1$ with probability $b$ and value $0$ with probability $1-b_i$, and $\lambda < 0$. Parameter $b$ controls how much of the vocabulary is forgotten and $\lambda$ is a soft penalty of including ``forgotten'' words in a review. 
$\lambda q_k$ emphasizes sentence forming with non-penalized words. The randomness is reset at the start of generating a new review.
Using Bernoulli penalties in the language model, we can ``forget'' a certain proportion of words and essentially ``force'' the creation of less typical sentences. We will test the effect of these two parameters, the Bernoulli probability $b$ and log-likelihood penalty of including ``forgotten'' words $\lambda$, with a user study in Section~\ref{sec:varying}.

\paragraph{Start penalty}
We introduce start penalties to avoid generic sentence starts (e.g. ``Great food, great service''). Inspired by \cite{li2016diversity}, we add a random start penalty $\lambda s^\mathrm{i}$, to our language model, which decreases monotonically for each generated token. We set $\alpha \leftarrow 0.66$ as it's effect decreases by 90\% every 5 words generated. 

\paragraph{Penalty for reusing words}
Bernoulli penalties do not prevent excessive use of certain words in a sentence (such as \textit{great} in Example~2).
To avoid excessive reuse of words, we included a memory penalty for previously used words in each translation. 
Concretely, we add the penalty $\lambda$ to each word that has been generated by the greedy search.

\subsubsection{Improving sentence coherence}
\label{sec:grammar}
We visually analyzed reviews after applying these penalties to our NMT model. While the models were clearly diverse, they were \emph{incoherent}: the introduction of random penalties had degraded the grammaticality of the sentences. Amongst others, the use of punctuation was erratic, and pronouns were used semantically wrongly (e.g. \emph{he}, \emph{she} might be replaced, as could ``and''/``but''). To improve the authenticity of our reviews, we added several \emph{grammar-based rules}. 

English language has several classes of words which are important for the natural flow of sentences. 
We built a list of common pronouns (e.g. I, them, our), conjunctions (e.g. and, thus, if), punctuation (e.g. ,/.,..), and apply only half memory penalties for these words. We found that this change made the reviews more coherent. The pseudocode for this and the previous step is shown in Algorithm~\ref{alg:aug}.
The combined effect of grammar-based rules and LM augmentation is visible in Example~3, Figure~\ref{fig:comparison}.

\begin{algorithm}[!t]
 \KwData{Initial log LM $\log p$, Bernoulli probability $b$, soft-penalty $\lambda$, monotonic factor $\alpha$, last generated token $o_i$, grammar rules set $G$}
 \KwResult{Augmented log LM $\log \Tilde{p}$}
\begin{algorithmic}[1]
\Procedure {Augment}{$\log p$, $b$, $\lambda$, $\alpha$, $o_i$, $i$}{ \\
generate $P_{\mathrm{1:N}} \leftarrow Bernoulli(b)$~~~~~~~~~~~~~~~|~$\text{One value} \in \{0,1\}~\text{per token}$~ \\
$I \leftarrow P>0$ ~~~~~~~~~~~~~~~~~~~~~~~~~~~~~~~~~~~~~~~|~Select positive indices~\\
$\log \Tilde{p} \leftarrow$ $\text{Discount}$($\log p$, $I$, $\lambda \cdot \alpha^i$,$G$) ~~~~~~ |~start penalty~\\ 
$\log \Tilde{p} \leftarrow$ $\text{Discount}$($\log \Tilde{p}$, $[o_i]$, $\lambda$,$G$) ~~~~~~~~~ |~memory penalty~\\ 
\textbf{return}~$\log \Tilde{p}$
}
\EndProcedure
\\
\Procedure {Discount}{$\log p$, $I$, $\lambda$, $G$}{
\State{\For{$i \in I$}{ 
\eIf{$o_i \in G$}{
$\log p_{i} \leftarrow \log p_{i} + \lambda/2$
}{
$\log p_{i} \leftarrow \log p_{i} + \lambda$}
}
\textbf{return}~$\log p$
\EndProcedure
}}
\end{algorithmic}
\caption{Pseudocode for augmenting language model. }
\label{alg:aug}
\end{algorithm}

\subsubsection{Human-like errors}
\label{sec:obfuscation}
We notice that our NMT model produces reviews without 
grammar mistakes. 
This is unlike real human writers, whose sentences contain two types of language mistakes 1) \emph{typos} that are caused by mistakes in the human motoric input, and 2) \emph{common spelling mistakes}.
We scraped a list of common English language spelling mistakes from Oxford dictionary\footnote{\url{https://en.oxforddictionaries.com/spelling/common-misspellings}} and created 80 rules for randomly \emph{re-introducing spelling mistakes}. 
Similarly, typos are randomly reintroduced based on the weighted edit distance\footnote{\url{https://pypi.python.org/pypi/weighted-levenshtein/0.1}}, such that typos resulting in real English words with small perturbations are emphasized. 
We use autocorrection tools\footnote{\url{https://pypi.python.org/pypi/autocorrect/0.1.0}} for finding these words.
We call these augmentations \emph{obfuscations}, since they aim to confound the reader to think a human has written them. We omit the pseudocode description for brevity.

\subsection{Experiment: Varying generation parameters in our NMT model}
\label{sec:varying}

Parameters $b$ and $\lambda$ control different aspects in fake reviews. 
We show six different examples of generated fake reviews in Table~\ref{table:categories}. 
Here, the largest differences occur with increasing values of $b$: visibly, the restaurant reviews become more extreme. 
This occurs because a large portion of vocabulary is ``forgotten''. Reviews with $b \geq 0.7$ contain more rare word combinations, e.g. ``!!!!!'' as punctuation, and they occasionally break grammaticality (''experience was awesome''). 
Reviews with lower $b$ are more generic: they contain safe word combinations like ``Great place, good service'' that occur in many reviews. Parameter $\lambda$'s is more subtle: it affects how random review starts are and to a degree, the discontinuation between statements within the review. 
We conducted an Amazon Mechanical Turk (MTurk) survey in order to determine what kind of NMT-Fake reviews are convincing to native English speakers. We describe the survey and results in the next section.

\begin{table}[!b]
\caption{Six different parametrizations of our NMT reviews and one example for each. The context is ``5 P~.~F~.~Chang ' s Scottsdale AZ'' in all examples.}
\begin{center}
  \begin{tabular}{ | l | l |  }
    \hline
    $(b, \lambda)$  & Example review for context \\ \hline
   \hline
   $(0.3, -3)$ & I love this location! Great service, great food and the best drinks in Scottsdale. \\
   & The staff is very friendly and always remembers u when we come in\\\hline
   $(0.3, -5)$ & Love love the food here! I always go for lunch. They have a great menu and \\
   & they make it fresh to order. Great place, good service and nice staff\\\hline
   $(0.5, -4)$ & I love their chicken lettuce wraps and fried rice!! The service is good, they are\\
   & always so polite. They have great happy hour specials and they have a lot\\
   & of options.\\\hline
   $(0.7, -3)$ & Great place to go with friends! They always make sure your dining \\
   & experience was awesome.\\ \hline
   $(0.7, -5)$ & Still haven't ordered an entree before but today we tried them once..\\ 
   & both of us love this restaurant....\\\hline
   $(0.9, -4)$ & AMAZING!!!!! Food was awesome with excellent service. Loved the lettuce \\
   & wraps. Great drinks and wine! Can't wait to go back so soon!!\\  \hline  
  \end{tabular}
  \label{table:categories}
\end{center}
\end{table}

\subsubsection{MTurk study}
\label{sec:amt}
We created 20 jobs, each with 100 questions, and requested master workers in MTurk to complete the jobs. 
We randomly generated each survey for the participants. Each review had a 50\% chance to be real or fake. The fake ones further were chosen among six (6) categories of fake reviews (Table~\ref{table:categories}). 
The restaurant and the city was given as contextual information to the participants. Our aim was to use this survey to understand how well English-speakers react to different parametrizations of NMT-Fake reviews. 
Table~\ref{table:amt_pop} in Appendix summarizes the statistics for respondents in the survey. All participants were native English speakers from America. The base rate (50\%) was revealed to the participants prior to the study.

We first investigated overall detection of any NMT-Fake reviews (1,006 fake reviews and 994 real reviews). We found that the participants had big difficulties in detecting our fake reviews. In average, the reviews were detected with class-averaged \emph{F-score of only 56\%}, with 53\% F-score for fake review detection and 59\% F-score for real review detection. The results are very close to \emph{random detection}, where precision, recall and F-score would each be 50\%. Results are recorded in Table~\ref{table:MTurk_super}. Overall, the fake review generation is very successful, since human detection rate across categories is close to random. 

\begin{table}[t]
\caption{Effectiveness of Mechanical Turkers in distinguishing human-written reviews from fake reviews generated by our NMT model (all variants).}
\begin{center}
  \begin{tabular}{ | c | c |c |c | c |  }
  \hline 
  \multicolumn{5}{|c|}{Classification report}
    \\ \hline 
    Review Type  & Precision & Recall & F-score & Support \\ \hline
    \hline
   Human & 55\% & 63\% & 59\% & 994\\
   NMT-Fake & 57\% & 50\% & 53\% & 1006 \\
   \hline
  \end{tabular}
  \label{table:MTurk_super}
\end{center}
\end{table}

We noticed some variation in the detection of different fake review categories. The respondents in our MTurk survey had most difficulties recognizing reviews of category $(b=0.3, \lambda=-5)$, where true positive rate was $40.4\%$, while the true negative rate of the real class was $62.7\%$. The precision were $16\%$ and $86\%$, respectively. The class-averaged F-score is $47.6\%$, which is close to random. Detailed classification reports are shown in Table~\ref{table:MTurk_sub} in Appendix. Our MTurk-study shows that \emph{our NMT-Fake reviews pose a significant threat to review systems}, since \emph{ordinary native English-speakers have very big difficulties in separating real reviews from fake reviews}. We use the review category $(b=0.3, \lambda=-5)$ for future user tests in this paper, since MTurk participants had most difficulties detecting these reviews. We refer to this category as NMT-Fake* in this paper.

\section{Evaluation}
\graphicspath{ {figures/}}

We evaluate our fake reviews by first comparing them statistically to previously proposed types of fake reviews, and proceed with a user study with experienced participants. We demonstrate the statistical difference to existing fake review types \cite{yao2017automated,mukherjee2013yelp,rayana2015collective} by training classifiers to detect previous types and investigate classification performance. 

\subsection{Replication of state-of-the-art model: LSTM}
\label{sec:repl}

Yao et al. \cite{yao2017automated} presented the current state-of-the-art generative model for fake reviews. The model is trained over the Yelp Challenge dataset using a two-layer character-based LSTM model. 
We requested the authors of \cite{yao2017automated} for access to their LSTM model or a fake review dataset generated by their model. Unfortunately they were not able to share either of these with us. 
We therefore replicated their model as closely as we could, based on their paper and e-mail correspondence\footnote{We are committed to sharing our code with bonafide researchers for the sake of reproducibility.}. 

We used the same graphics card (GeForce GTX) and trained using the same framework (torch-RNN in lua). We downloaded the reviews from Yelp Challenge and preprocessed the data to only contain printable ASCII characters, and filtered out non-restaurant reviews. We trained the model for approximately 72 hours. We post-processed the reviews using the customization methodology described in \cite{yao2017automated} and email correspondence. We call fake reviews generated by this model LSTM-Fake reviews.

\subsection{Similarity to existing fake reviews}
\label{sec:automated}

We now want to understand how NMT-Fake* reviews compare to a) LSTM fake reviews and b) human-generated fake reviews. We do this by comparing the statistical similarity between these classes. 

\begin{figure}
  \begin{subfigure}[b]{0.5\columnwidth}
    \includegraphics[width=\columnwidth]{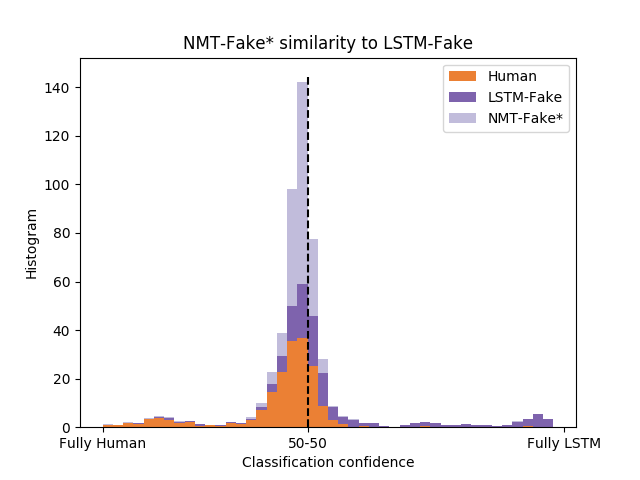}
    \caption{Human--LSTM reviews.}
    \label{fig:lstm}
  \end{subfigure}
  \begin{subfigure}[b]{0.5\columnwidth}
    \includegraphics[width=\columnwidth]{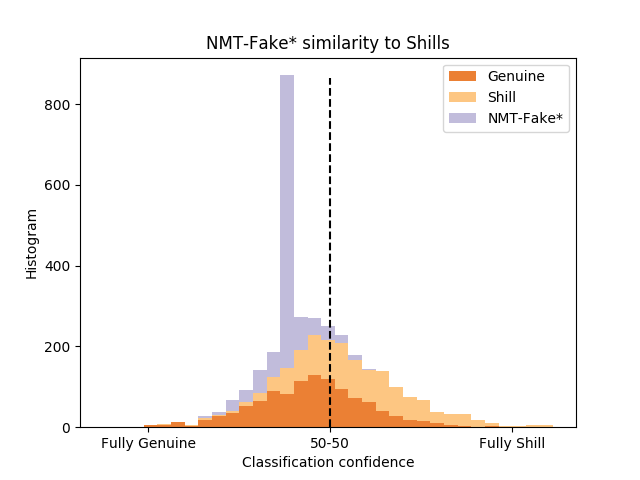}
    \caption{Genuine--Shill reviews.}
    \label{fig:shill}
  \end{subfigure}
  \caption{
  Histogram comparison of NMT-Fake* reviews with LSTM-Fake reviews and human-generated (\emph{genuine} and \emph{shill}) reviews. Figure~\ref{fig:lstm} shows that a classifier trained to distinguish ``human'' vs. LSTM-Fake cannot distinguish ``human'' vs NMT-Fake* reviews. Figure~\ref{fig:shill} shows NMT-Fake* reviews are more similar to \emph{genuine} reviews than \emph{shill} reviews.
  }
  \label{fig:statistical_similarity}
\end{figure}

For `a' (Figure~\ref{fig:lstm}), we use the Yelp Challenge dataset. We trained a classifier using 5,000 random reviews from the Yelp Challenge dataset (``human'') and 5,000 fake reviews generated by LSTM-Fake. Yao et al. \cite{yao2017automated} found that character features are essential in identifying LSTM-Fake reviews. Consequently, we use character features (n-grams up to 3).

For `b' (Figure~\ref{fig:shill}),we the ``Yelp Shills'' dataset (combination of YelpZip \cite{mukherjee2013yelp}, YelpNYC \cite{mukherjee2013yelp}, YelpChi \cite{rayana2015collective}). This dataset labels entries that are identified as fraudulent by Yelp's filtering mechanism (''shill reviews'')\footnote{Note that shill reviews are probably generated by human shills \cite{zhao2017news}.}. The rest are treated as genuine reviews from human users (''genuine''). We use 100,000 reviews from each category to train a classifier. We use features from the commercial psychometric tool LIWC2015 \cite{pennebaker2015development} to generated features.

In both cases, we use AdaBoost (with 200 shallow decision trees) for training. For testing each classifier, we use a held out test set of 1,000 reviews from both classes in each case. In addition, we test 1,000 NMT-Fake* reviews. Figures~\ref{fig:lstm} and~\ref{fig:shill} show the results. The classification threshold of 50\% is marked with a dashed line.

We can see that our new generated reviews do not share strong attributes with previous known categories of fake reviews. If anything, our fake reviews are more similar to genuine reviews than previous fake reviews. We thus conjecture that our NMT-Fake* fake reviews present a category of fake reviews that may go undetected on online review sites.

\subsection{Comparative user study}
\label{sec:comparison}
We wanted to evaluate the effectiveness of fake reviews againsttech-savvy users who understand and know to expect machine-generated fake reviews. We conducted a user study with 20 participants, all with computer science education and at least one university degree. Participant demographics are shown in Table~\ref{table:amt_pop} in the Appendix. Each participant first attended a training session where they were asked to label reviews (fake and genuine) and could later compare them to the correct answers -- we call these participants \emph{experienced participants}.
No personal data was collected during the user study.

Each person was given two randomly selected sets of 30 of reviews (a total of 60 reviews per person) with reviews containing 10 \textendash 50 words each. 
Each set contained 26 (87\%) real reviews from Yelp and 4 (13\%) machine-generated reviews,
numbers chosen based on suspicious review prevalence on Yelp~\cite{mukherjee2013yelp,rayana2015collective}. 
One set contained machine-generated reviews from one of the two models (NMT ($b=0.3, \lambda=-5$) or LSTM), 
and the other set reviews from the other in randomized order. 
The number of fake reviews was revealed to each participant in the study description. Each participant was requested to mark four (4) reviews as fake.

Each review targeted a real restaurant. A screenshot of that restaurant's Yelp page was shown to each participant prior to the study. Each participant evaluated reviews for one specific, randomly selected, restaurant. An example of the first page of the user study is shown in Figure~\ref{fig:screenshot} in Appendix.

\begin{figure}[!ht]
\centering
\includegraphics[width=.7\columnwidth]{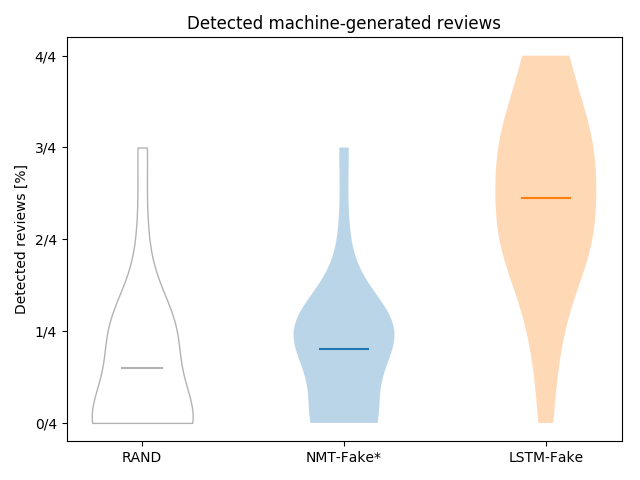}
\caption{Violin plots of detection rate in comparative study. Mean and standard deviations for number of detected fakes are $0.8\pm0.7$ for NMT-Fake* and $2.5\pm1.0$ for LSTM-Fake. $n=20$. A sample of random detection is shown as comparison.}
\label{fig:aalto}
\end{figure}

Figure~\ref{fig:aalto} shows the distribution of detected reviews of both types. A hypothetical random detector is shown for comparison.
NMT-Fake* reviews are significantly more difficult to detect for our experienced participants. In average, detection rate (recall) is $20\%$ for NMT-Fake* reviews, compared to $61\%$ for LSTM-based reviews. 
The precision (and F-score) is the same as the recall in our study, since participants labeled 4 fakes in each set of 30 reviews \cite{murphy2012machine}. 
The distribution of the detection across participants is shown in Figure~\ref{fig:aalto}. \emph{The difference is statistically significant with confidence level $99\%$} (Welch's t-test). 
We compared the detection rate of NMT-Fake* reviews to a random detector, 
and find that \emph{our participants detection rate of NMT-Fake* reviews is not statistically different from random predictions with 95\% confidence level} (Welch's t-test).

\section{Defenses}

\label{sec:detection}

We developed an AdaBoost-based classifier to detect our new fake reviews, consisting of 200 shallow decision trees (depth 2). The features we used are recorded in Table~\ref{table:features_adaboost} (Appendix).  
We used word-level features based on spaCy-tokenization \cite{honnibal-johnson:2015:EMNLP} and constructed n-gram representation of POS-tags and dependency tree tags. We added readability features from NLTK~\cite{bird2004nltk}.

\begin{figure}[ht]
\centering
\includegraphics[width=.7\columnwidth]{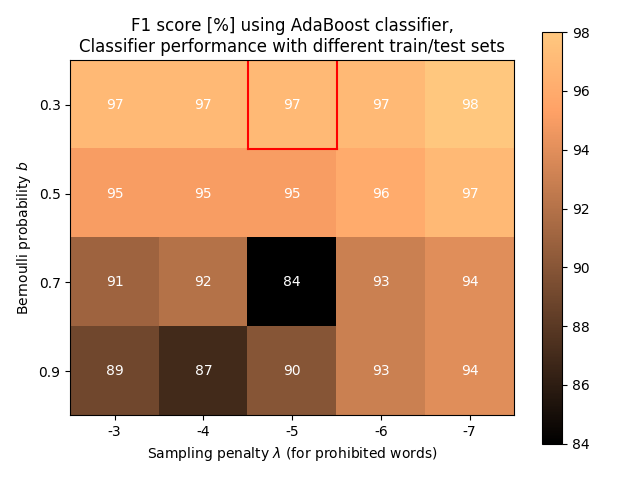}
\caption{
Adaboost-based classification of NMT-Fake and human-written reviews. 
Effect of varying $b$ and $\lambda$ in fake review generation. 
The variant native speakers had most difficulties detecting is well detectable by AdaBoost (97\%).}
\label{fig:adaboost_matrix_b_lambda}
\end{figure}

Figure~\ref{fig:adaboost_matrix_b_lambda} shows our AdaBoost classifier's class-averaged F-score at detecting different kind of fake reviews. The classifier is very effective in detecting reviews that humans have difficulties detecting. For example, the fake reviews MTurk users had most difficulty detecting ($b=0.3, \lambda=-5$) are detected with an excellent 97\% F-score. 
The most important features for the classification were counts for frequently occurring words in fake reviews (such as punctuation, pronouns, articles) as well as the readability feature ``Automated Readability Index''. We thus conclude that while NMT-Fake reviews are difficult to detect for humans, they can be well detected with the right tools. 

\section{Related Work}

Kumar and Shah~\cite{kumar2018false} survey and categorize false information research. Automatically generated fake reviews are a form of \emph{opinion-based false information}, where the creator of the review may influence reader's opinions or decisions. 
Yao et al. \cite{yao2017automated} presented their study on machine-generated fake reviews. Contrary to us, they investigated character-level language models, without specifying a specific context before generation. We leverage existing NMT tools to encode a specific context to the restaurant before generating reviews. 
Supporting our study, Everett et al~\cite{Everett2016Automated} found that security researchers were less likely to be fooled by Markov chain-generated Reddit comments compared to ordinary Internet users.

Diversification of NMT model outputs has been studied in \cite{li2016diversity}. The authors proposed the use of a penalty to commonly occurring sentences (\emph{n-grams}) in order to emphasize maximum mutual information-based generation. 
The authors investigated the use of NMT models in chatbot systems. 
We found that unigram penalties to random tokens (Algorithm~\ref{alg:aug}) was easy to implement and produced sufficiently diverse responses.

\section {Discussion and Future Work}

\paragraph{What makes NMT-Fake* reviews difficult to detect?} First, NMT models allow the encoding of a relevant context for each review, which narrows down the possible choices of words that the model has to choose from. Our NMT model had a perplexity of approximately $25$, while the model of \cite{yao2017automated} had a perplexity of approximately $90$ \footnote{Personal communication with the authors}. Second, the beam search in NMT models narrows down choices to natural-looking sentences. Third, we observed that the NMT model produced \emph{better structure} in the generated sentences (i.e. a more coherent story).

\paragraph{Cost of generating reviews} With our setup, generating one review took less than one second. The cost of generation stems mainly from the overnight training. Assuming an electricity cost of 16 cents / kWh (California) and 8 hours of training,  training the NMT model requires approximately 1.30 USD. This is a 90\% reduction in time compared to the state-of-the-art \cite{yao2017automated}. Furthermore, it is possible to generate both positive and negative reviews with the same model.

\paragraph{Ease of customization} We experimented with inserting specific words into the text by increasing their log likelihoods in the beam search. We noticed that the success depended on the prevalence of the word in the training set. For example, adding a +5 to \emph{Mike} in the log-likelihood resulted in approximately 10\% prevalence of this word in the reviews. An attacker can therefore easily insert specific keywords to reviews, which can increase evasion probability.

\paragraph{Ease of testing} Our diversification scheme is applicable during \emph{generation phase}, and does not affect the training setup of the network in any way. Once the NMT model is obtained, it is easy to obtain several different variants of NMT-Fake reviews by varying parameters $b$ and $\lambda$. 



\paragraph{Languages} The generation methodology is not per-se language-dependent. The requirement for successful generation is that sufficiently much data exists in the targeted language. However, our language model modifications require some knowledge of that target language's grammar to produce high-quality reviews.

\paragraph{Generalizability of detection techniques} Currently, fake reviews are not universally detectable. Our results highlight that it is difficult to claim detection performance on unseen types of fake reviews (Section~\ref{sec:automated}). We see this an open problem that deserves more attention in fake reviews research.

\paragraph{Generalizability to other types of datasets} Our technique can be applied to any dataset, as long as there is sufficient training data for the NMT model. We used approximately 2.9 million reviews for this work.

\section{Conclusion}

In this paper, we showed that neural machine translation models can be used to generate fake reviews that are very effective in deceiving even experienced, tech-savvy users. 
This supports anecdotal evidence \cite{national2017commission}.
Our technique is more effective than state-of-the-art \cite{yao2017automated}. 
We conclude that machine-aided fake review detection is necessary since human users are ineffective in identifying fake reviews. 
We also showed that detectors trained using one type of fake reviews are not effective in identifying other types of fake reviews.
Robust detection of fake reviews is thus still an open problem.

\section*{Acknowledgments}
We thank Tommi Gr\"{o}ndahl for assistance in planning user studies and the 
participants of the user study for their time and feedback. We also thank 
Luiza Sayfullina for comments that improved the manuscript. 
We thank the authors of \cite{yao2017automated} for answering questions about 
their work.

\bibliographystyle{splncs}
\bibliography{ref}

\section*{Appendix}

We present basic demographics of our MTurk study and the comparative study with experienced users in Table~\ref{table:amt_pop}.

\begin{table}
\caption{User study statistics.}
\begin{center}
  \begin{tabular}{ | l | c | c |  }
    \hline
    Quality & Mechanical Turk users & Experienced users\\
    \hline
    Native English Speaker & Yes (20) & Yes (1) No (19) \\
    Fluent in English & Yes (20) & Yes (20) \\
    Age & 21-40 (17) 41-60 (3) & 21-25 (8) 26-30 (7) 31-35 (4) 41-45 (1)\\
    Gender & Male (14) Female (6) & Male (17) Female (3)\\
    Highest Education & High School (10) Bachelor (10) &  Bachelor (9) Master (6) Ph.D. (5) \\
   \hline
  \end{tabular}
  \label{table:amt_pop}
\end{center}
\end{table}

Table~\ref{table:openNMT-py_commands} shows a listing of the openNMT-py commands we used to create our NMT model and to generate fake reviews.

\begin{table}[t]
\caption{Listing of used openNMT-py commands.}
\begin{center}
  \begin{tabular}{ | l | l |  }
    \hline
    Phase & Bash command \\
    \hline
   Preprocessing & \begin{lstlisting}[language=bash]
python preprocess.py -train_src context-train.txt
-train_tgt reviews-train.txt -valid_src context-val.txt
-valid_tgt reviews-val.txt -save_data model
-lower -tgt_words_min_frequency 10
\end{lstlisting}
  \\ & \\
   Training &  \begin{lstlisting}[language=bash]
python train.py -data model -save_model model -epochs 8 
-gpuid 0 -learning_rate_decay 0.5 -optim adam
-learning_rate 0.001 -start_decay_at 3\end{lstlisting}
  \\ & \\
   Generation &  \begin{lstlisting}[language=bash]
python translate.py -model model_acc_35.54_ppl_25.68_e8.pt
-src context-tst.txt -output pred-e8.txt -replace_unk
-verbose -max_length 50 -gpu 0
   \end{lstlisting} \\
  \hline
  \end{tabular}
  \label{table:openNMT-py_commands}
\end{center}
\end{table}


Table~\ref{table:MTurk_sub} shows the classification performance of Amazon Mechanical Turkers, separated across different categories of NMT-Fake reviews. The category with best performance ($b=0.3, \lambda=-5$) is denoted as NMT-Fake*.

\begin{table}[b]
\caption{MTurk study subclass classification reports. Classes are imbalanced in ratio 1:6. Random predictions are $p_\mathrm{human} = 86\%$ and $p_\mathrm{machine} = 14\%$, with $r_\mathrm{human} = r_\mathrm{machine} = 50\%$. Class-averaged F-scores for random predictions are $42\%$.}
\begin{center}
  \begin{tabular}{ | c || c |c |c | c |  }
    \hline
    $(b=0.3, \lambda = -3)$  & Precision & Recall & F-score & Support \\ \hline
   Human & 89\% & 63\% & 73\% & 994\\
   NMT-Fake & 15\% & 45\% & 22\% & 146 \\
   \hline
    \hline
    $(b=0.3, \lambda = -5)$  & Precision & Recall & F-score & Support \\ \hline
   Human & 86\% & 63\% & 73\% & 994\\
   NMT-Fake* & 16\% & 40\% & 23\% & 171 \\
   \hline
    \hline
    $(b=0.5, \lambda = -4)$  & Precision & Recall & F-score & Support \\ \hline
   Human & 88\% & 63\% & 73\% & 994\\
   NMT-Fake & 21\% & 55\% & 30\% & 181 \\
   \hline
    \hline
    $(b=0.7, \lambda = -3)$  & Precision & Recall & F-score & Support \\ \hline
   Human & 88\% & 63\% & 73\% & 994\\
   NMT-Fake & 19\% & 50\% & 27\% & 170 \\
   \hline
    \hline
    $(b=0.7, \lambda = -5)$  & Precision & Recall & F-score & Support \\ \hline
   Human & 89\% & 63\% & 74\% & 994\\
   NMT-Fake & 21\% & 57\% & 31\% & 174 \\
   \hline
    \hline
    $(b=0.9, \lambda = -4)$  & Precision & Recall & F-score & Support \\ \hline
   Human & 88\% & 63\% & 73\% & 994\\
   NMT-Fake & 18\% & 50\% & 27\% & 164 \\
   \hline
  \end{tabular}
  \label{table:MTurk_sub}
\end{center}
\end{table}

\begin{figure}[ht]
\centering
\includegraphics[width=1.\columnwidth]{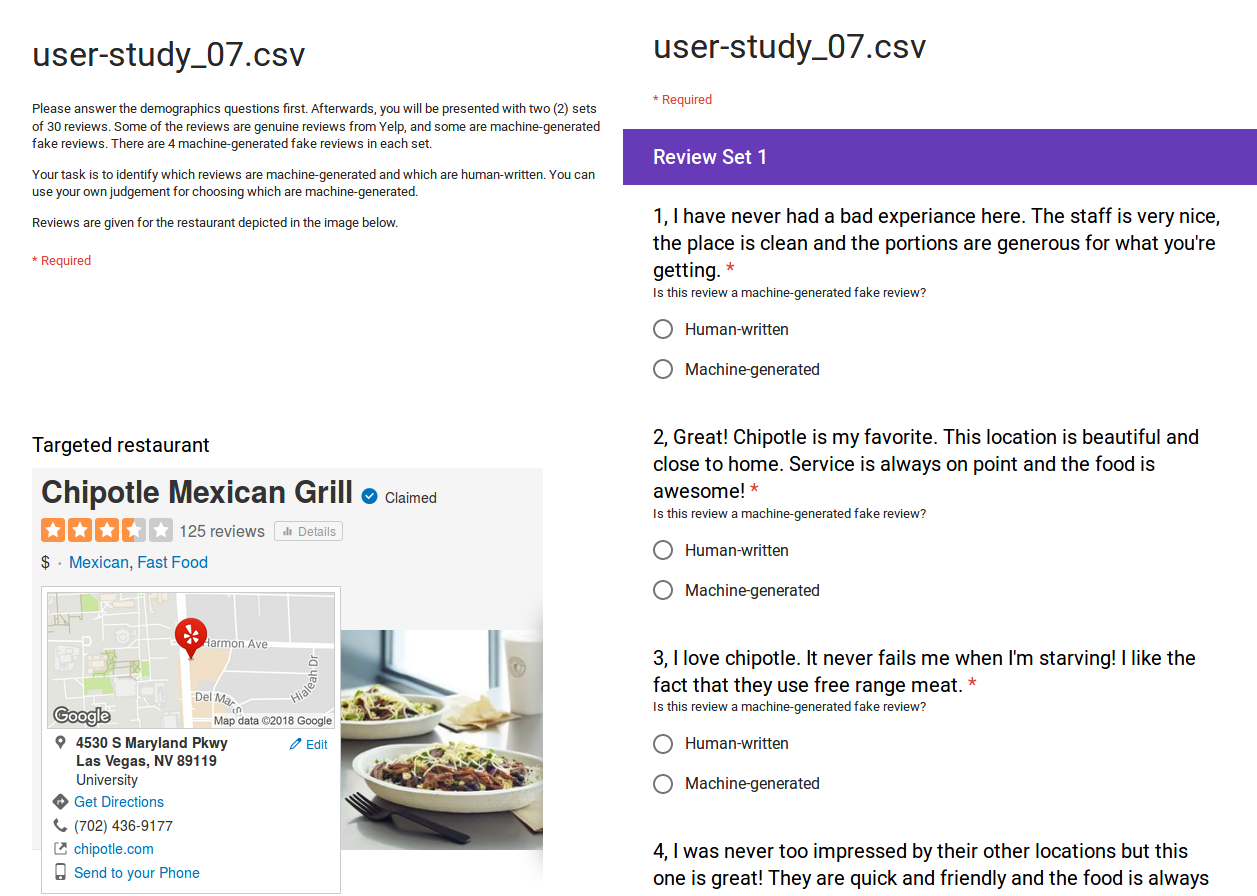}
\caption{
Screenshots of the first two pages in the user study. Example 1 is a NMT-Fake* review, the rest are human-written.
}
\label{fig:screenshot}
\end{figure}

Figure~\ref{fig:screenshot} shows screenshots of the first two pages of our user study with experienced participants.

Table~\ref{table:features_adaboost} shows the features used to detect NMT-Fake reviews using the AdaBoost classifier.

\begin{table}
\caption{Features used in NMT-Fake review detector.}
\begin{center}
  \begin{tabular}{ | l | c |  }
    \hline
    Feature type & Number of features \\ \hline
    \hline
   Readability features & 13  \\ \hline
   Unique POS tags & $~20$  \\ \hline
   Word unigrams & 22,831  \\ \hline
   1/2/3/4-grams of simple part-of-speech tags & 54,240  \\ \hline
   1/2/3-grams of detailed part-of-speech tags & 112,944  \\ \hline   
   1/2/3-grams of syntactic dependency tags & 93,195  \\ \hline   
  \end{tabular}
  \label{table:features_adaboost}
\end{center}
\end{table}

\end{document}